\documentclass[aip,prb,amsmath,amssymb,reprint]{revtex4}
\usepackage[dvips]{graphicx}
\usepackage{dcolumn}
\usepackage{color}
\usepackage{ulem}
\usepackage{mathptmx, helvet, courier}
\usepackage{threeparttable}
\usepackage{multirow}

\begin{document}

\title{Structural entanglements in protein complexes}

\author{Yani Zhao}
\affiliation{Institute of Physics, Polish Academy of Sciences,
Aleja Lotnik{\'o}w 32/46, PL-02668 Warsaw, Poland}

\author{Mateusz Chwastyk}
\affiliation{Institute of Physics, Polish Academy of Sciences,
Aleja Lotnik{\'o}w 32/46, PL-02668 Warsaw, Poland}

\author{Marek Cieplak}
\email{mc@ifpan.edu.pl}
\affiliation{Institute of Physics, Polish Academy of Sciences,
Aleja Lotnik{\'o}w 32/46, PL-02668 Warsaw, Poland}

\begin{abstract}
We consider multi-chain protein native structures and propose a criterion
that determines whether two chains in the system are entangled or not.
The criterion is based on the behavior observed by pulling at both
temini of each chain simultaneously in the two chains. We have identified
about 900 entangled systems in the Protein Data Bank and provided a more detailed
analysis for several of them. We argue that entanglement enhances the
thermodynamic stability of the system but it may have other functions:
burying the hydrophobic residues at the interface,
and increasing the DNA or RNA binding area.
We also study the folding and stretching properties of the knotted dimeric
proteins MJ0366, YibK and bacteriophytochrome. These proteins have been
 studied theoretically     
in their monomeric versions so far. The dimers are seen to separate
on stretching through the tensile mechanism and the characteristic unraveling
force depends on the pulling direction.
\end{abstract}

\maketitle

\section{Introduction}

There is a growing interest in proteins that contain knots
\cite{Virnau,Virnau1,Stasiak}. Chronologically, the first entry on the list of
such proteins appears to be carbonic anhydrase B \cite{Richardson,Mansfield}
and the current status of the list
can be obtained, for instance, at the KnotProt database \cite{KnotProt}.
The protein backbones do not form closed loops, however, and the presence
of a knot cannot be assessed in absolute terms as it depends on the
method used to connect the termini beyond the backbone in order to
generate a closed line. A common computational approach to determine
the location of a knot is to use the KMT
algorithm \cite{km1,taylor} which reduces the topological complexity of
the backbone in a step-wise fashion by decreasing the number of segments.
The reduced structure brings out the topology and allows for
a determination of the sequential locations of the knot's ends.
It should be noted that
this approach does not impose any closure of the backbone line
which may sometimes result in not detecting a knot that appears
to be present. The KMT algorithm leads to an observation that 
carbonic anhydrase B contains a shallow knot, since
one of the knot ends is just three sites away from the C-terminus.
Topologically, it is very easy to untie it and then tie it back.
On the other hand, for deeply knotted proteins, both ends of the knot are at a
substantial distance away from the termini and tying such a knot is much harder.
An experimental way to tell the existence of the knot is to stretch a monomeric protein
by the terminal sites \cite{Rief,Sulkowska_2008,Dziubiella,Israel}.
The knot is thought to be present if the utmost backbone
extension is noticeably shorter than the contour length, as the
tightened knot takes away a portion of the chain that could be extended.

\vspace*{0.5cm}

In this paper, we consider multi-chain protein complexes and
focus on the entanglement effects that arise from interactions
between separate unknotted chains. In some complexes, the
individual chains may themselves be entangled when they contain knots.
In fact, many of the well studied knotted proteins,
such as the deeply knotted YibK methyltransferase from
{\it Haemophilus influenzae} (PDB:1J85) \cite{1J85,nureki}
and the shallowly knotted  MJ0366 from  methanogenic
archea {\it Methanocaldococcus jannaschi} (PDB:2EFV) are actually
dimeric.  In these systems, the native entanglement is found in
each of the chains individually and, in principle, can be attested
experimentally through stretching of single chains.
However, such single-chain stretching does not test
the entanglements that arise from interactions between
two chains (knotted or unknotted).
We propose that the way to detect them
is by pulling at four termini instead of two. If the termini
of one chain are denoted by N and C, and of the other by N' and C',
then we propose to  pull by N--C and, simultaneously, by N'--C'.
Alternatively, we can anchor N and C while pulling simultaneously by N'--C'
 along the line connecting N' with C'.
The two chains are considered to be untangled if this action
results in separation of the chains. Otherwise, we declare
the chains to be entangled.  Sometimes, 
anchoring N' and C' combined with pulling by N and C may 
result in a different verdict. We refer to such situations
as type-I entanglement in distinction to type-II entanglement
which is unambiguous. We discuss this point later.

\vspace*{0.5cm}

It should be noted that the four-terminal pulling process of two chains
is quite distinct from the two-terminal stretching of a dimer.
 In the former case, pulling takes place in each of the chains
and the termini of each chain are pulled along the same direction. 
In the latter case, two termini are anchored and stretching 
takes place either in one chain or it works to separate
one chain from another in a variety of possible ways.
The chosen termini are then stretched in opposite directions. 
In a dimer,
there are six choices
to pick two termini out of four: N--C, N'--C', N--C',
N'--C, C--C', and N--N'.  Stretching is assumed to take
place along the line connecting the selected termini. 
The outcomes of the two-terminal processes
usually depend on this choice,  as demonstrated in the
context of unknotted proteins \cite{Sikoramdom,SikoraPRL,Miod,cohdoc}:
the chains may or may not separate on  stretching and the characteristic
unravelling forces are anisotropic. 
In the examples of knotted dimers considered here, 
the two chains are connected 
by a number of contacts but are not mutually entangled.
In these situations, the dynamics of the N--C  stretching is found to
be affected by the presence of the other chain only in a minor way,
but the cross-chain stretchings are very distinct.

\vspace*{0.5cm}

Here, we report on
 many instances of inter-chain entanglement found in
our survey of 10 498 $n$-meric protein structures,
where $n\ge 2$. The structures, both homomeric and heteromeric, were
obtained from the Protein Data Bank (PDB), in an essentially random way,
by considering all 4-character alphanumerical structure codes
in which the two middle characters are between "00" and "cz".
About 900 of these, i.e. 8.6\%, were found to be entangled.
In addition, we also consider six large ribosomal complexes, corresponding
to the PDB codes of 3JCT, 4V4J, 5FCJ, 5IT7, 5J88, and 5J86. 
We analyze a number of examples of the resulting
cases of entanglement in detail.
Our four-terminal pulling is implemented by using a structure-based
coarse-grained model \cite{biophysical,JPCM,models,PLOS}.
Any other molecular dynamics approach is expected to lead
to similar conclusions regarding the detection of entanglement.
However, the dynamics of the pulling process itself should depend on the
model and the pulling speed. As to the function of the interchain entanglement,
it is possible that it leads to an increased thermal stability of the system.

\vspace*{0.5cm}

Experiments on knotted proteins have been done for monomeric systems
such as ubuiquitin hydrolases UCH-L1 (PDB:2ETL)
\cite{ZhangJackson} and UCH-L3 (PDB:1XD1) \cite{AndersonJackson} or
a bacteriophotochrome (PDB:2O9C) \cite{Rief}, even though the
latter protein forms dimers in a solution.
(From now on, we shall refer to the proteins by their PDB
structure codes for brevity.)
Another purpose of this paper is to reconsider the
folding and stretching properties of the knotted
1J85, 2EFV, and 2O9C proteins by noticing the fact that they function
as dimers that are bound by hydrogen and hydrophobic bonds.
Unlike most of the cysteine-knot dimers \cite{SikoraPRL}, there are
no covalent bonds between the chains.
So far, theoretical studies of these systems have
been restricted to monomers. Monomeric models of 1J85 have been considered,
for instance, in refs. \cite{Sulkowska_2008,dodging,Israel,nascent} and
of 2EFV in refs. \cite{Micheletti,Noel,shallow}.
One of the results obtained was that the probability to form a knot
on folding is enhanced by on-ribosome folding \cite{nascent,shallow}.
Here, however, we do not consider folding under nascent conditions.
The three knotted dimers considered here are not entangled
when using the 4-termini pulling test -- the chains just separate.

\vspace*{0.5cm}

It should be noted that entanglements are known to arise and disappear
spontaneously in dense polymeric liquids \cite{McLeish,Everaers},
whereas the topological effects discussed here appear in the native
state of several polypeptide chains. Such chains have been considered
recently by Baiesi et al. \cite{Baiesi}. Specifically, they have
identified 110 domain-swapped dimers \cite{Benkovic,Jaskolski}
and determined their linking numbers \cite{Buck}.
These numbers, denoted here as $L_k$, are obtained by
closing the chains into loops and then evaluating the Gauss double
linking integrals. The $L_k$ indicates the number of times
that one closed curve winds around another  in the three-dimensional
space. Generally, a pair of chains
is entangled if the absolute value of $L_k$ is larger than 0
(one exception is the Whitehead link).

\vspace*{0.5cm}

The protein systems we study need not involve
domain swapping and need not have sequentially identical chains.
Most of them (over 85\%) are not domain-swapped but only a few are
sequentially heterogeneous. The list of the protein studied is
shown in Table~SI in the Supplementary Material (SM).

\section{Structure-based modeling}

We use a Go-like model \cite{Go0} with the specific implementation as
described in refs. \cite{JPCM, models, PLOS}. The length-related parameters
in the bonding and non-bonding potentials are derived from the native
structure. The molecular dynamics employed here deals only with the $\alpha$-C
atoms and the solvent is implicit.
The bonded interactions are described by the harmonic potentials.
Non-bonded interactions, or contacts, are assigned to pairs
of amino acids by using the overlap criterion in which
the heavy atoms in the native conformation are represented by enlarged
van der Waals spheres \cite{Tsai,Settanni,JPCM}. The contact exists if
at least two such spheres from different residues overlap.
These contacts are described by the Lennard-Jones potentials with the
minima at the crystallographically determined distances. The contact potentials
have a depth,  $\epsilon$, which is identical in each contact.
Non-native contacts are considered repulsive.
Variants with non-uniform values of $\epsilon$ have been shown \cite{models}
to yield statistically similar behavior in a test set of 28 proteins.

\vspace*{0.5cm}

The backbone stiffness is accounted for by the chirality potential \cite{JPCM}
which favors the native sense of the local backbone chirality.
The value of $\epsilon$
has been calibrated by making comparisons to the experimental
data on stretching: approximately, $\epsilon$/{\AA} is 110 pN
(which also is close to the energy of the O-H-N hydrogen bond
of 1.65 kcal/mol and close to 1.5 kcal/mol derived as a typical
value in loop parts of a protein through all-atom simulations \cite{Poma}).
For most unknotted proteins, optimal folding takes place around
the temperature, $T$, of 0.3~$\epsilon/k_B$
($k_B$ is the Boltzmann constant; the stiffness parameters depend on $\epsilon$)
and this temperature should correspond to a vicinity of the room $T$.

\vspace*{0.5cm}

We use the Langevin thermostat with substantial damping. The time unit
of the simulations, $\tau$, is effectively of order 1 ns as the displacement
of the atoms is dominated by diffusion instead of ballistic motions.
Folding is declared as accomplished when all native contacts are established
for the first time (the distance between two $\alpha$-C in a contact
is smaller than the native distance multiplied by 1.5).
For knotted proteins, however, this condition does not necessarily
signify formation of the native knot.
The situation in which there is no knot but all contacts are established
is referred to as misfolding.
It should be noted that the contacts can be set and then broken multiply due 
 to
the thermal fluctuations.

\vspace*{0.5cm}

When studying folding, we prepare the starting state of a dimeric protein
by heating the native structure to $T=1.5 \; \epsilon/k_B$ and make sure
that all of its native contacts are broken. Each folding trajectory,
at a much lower $T$, starts from a different conformation. Proper folding
is declared to take place if all native contacts are established
and so are the knots.
 The simulations are performed in an infinite space so some
of the trajectories may result in the two chains never meeting again.
Similar strategy is used to fold 2O9C monomers because of its covalently
bound ligand. In the holoprotein form (a conjugated protein contains a
cofactor), this ligand is not cleaved off from the initial conformation
during folding to prevent the situation that it fluctuates too much
to make a contact with the protein. In case when the ligand is reduced,
one obtains the apoprotein of 2O9C.
In the coarse-grained model, we represent the ligand of 2O9C by
20 effective atoms located at the backbone carbon atoms \cite{crys2o9c}.

\vspace*{0.5cm}

Pulling is accomplished at a constant speed, $v_p$, along the direction
that connects two termini in the native state.
 It has to be noted that, this choice of the direction is 
just one of many and the outcome of the procedure generally depends 
on it (see Fig. S1 in SM).
When determining
the existence of an entanglement, we use $v_p$ of 0.01~{\AA}/$\tau$.
The choice of the value of $v_p$ may affect the dynamics of the
testing process but has no relevance for the outcome regarding the
overall topology.
The termini involved in pulling are attached to harmonic springs of elastic
constant $k=0.12$~$\epsilon/${\AA}, which is close to the values
corresponding to the elasticity of cantilevers in atomic force microscopes.
For a given multi-chain protein, we consider all possible pair permutations.
The chains are considered to be separated if the centers
of mass of the chains are further away from each other than the
contour length of one chain. 
Entanglement shows as a rapid growth in the resisting force and we stop
the test if the force exceeds 25~$\epsilon$/{\AA}. When
studying the very process of stretching, we take $v_p$ of 0.005~{\AA}/$\tau$,
which is some 2 orders of magnitudes faster than typical experimental speeds.

\section{Interchain entanglements} \label{epair}

\begin{figure}[h]
\centering
\includegraphics[width=0.5\textwidth]{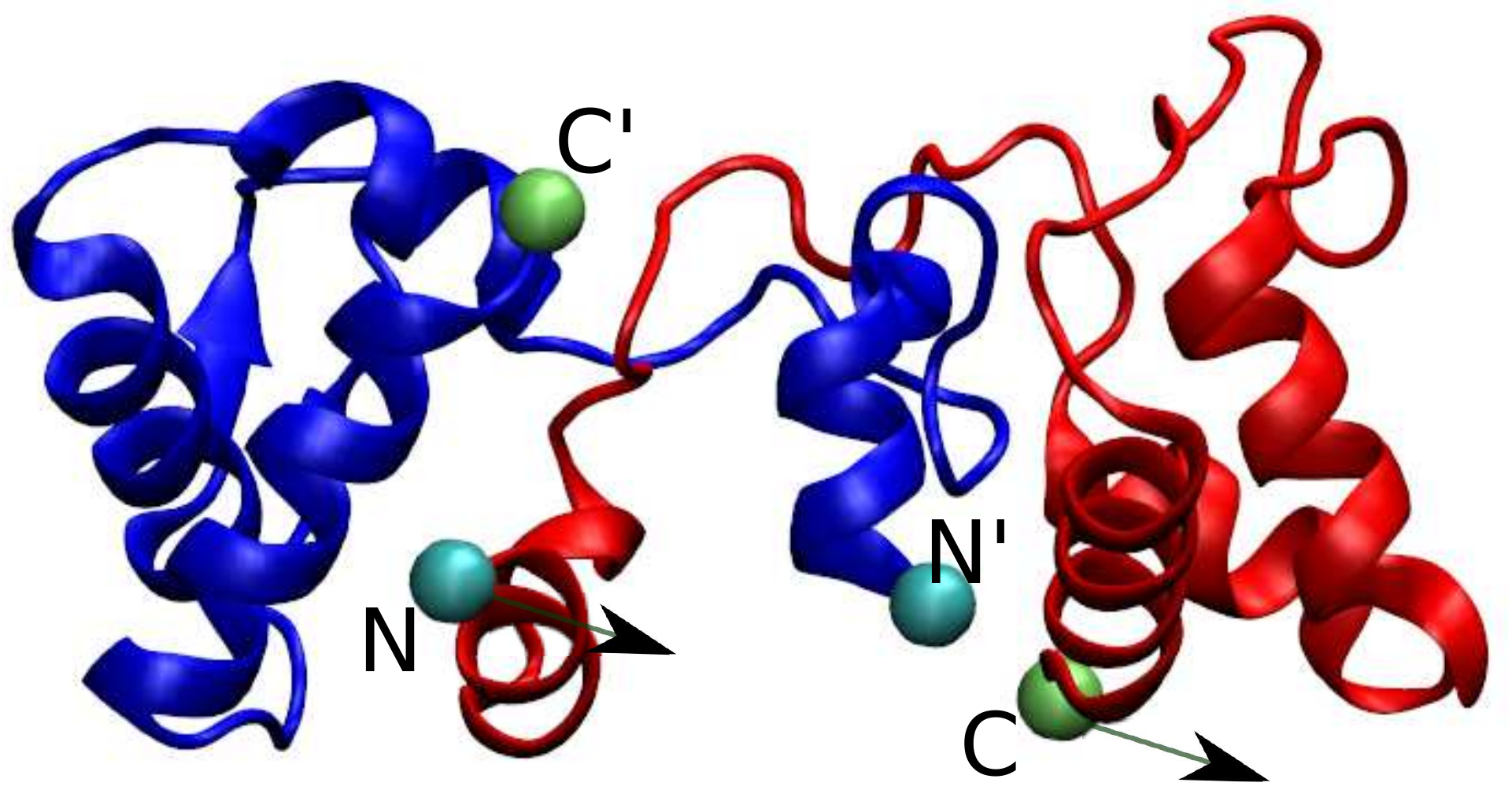}
\caption{An example of the direction of stretching for 5AUR. One chain (A) is
in red and another (C) in blue. The N and N' termini
are marked in cyan and the C and C' termini in lime.  The indicated 
direction of pulling leads to chains A and C being  entangled. }\label{examp5aur}
\end{figure}

\begin{figure}[h]
\centering
\includegraphics[width=0.5\textwidth]{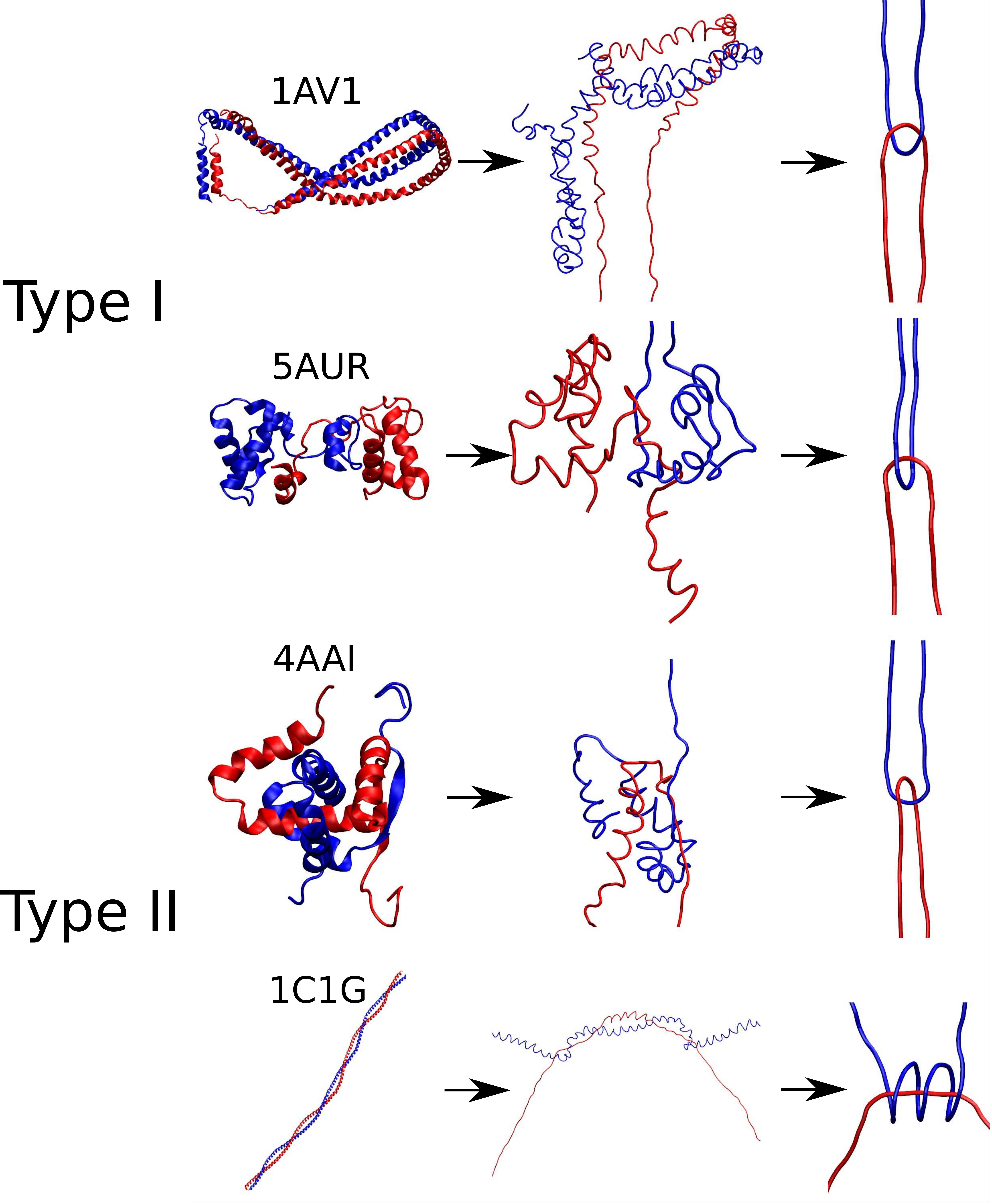}
\caption{  Stages of pulling at the termini N and 
C while anchoring  N' and C' simultaneously, or vice versa. 
 The top two examples (chains A and B
in 1AV1, chains A and C in 5AUR) correspond to entanglement of type I
and the remaining (chains A and B in 4AAI and 1C1G) -- of type II. The absolute value of $L_k$ of 1AV1, 5AUR, 4AAI and 1C1G is 1, 1, 1 and 3, respectively.
}\label{selectpdb}
\end{figure}

The self-association of proteins resulting in the formation of multimers is
a common phenomenon in nature. For example,
only a third of human enzymes are monomers, the rest are multimers
\cite{Matthews2004}. In bacteria ({\it Escherichia coli}), about 19.4\%
of proteins are monomeric and 38.2\% dimeric, the other are multimers composed of more than 2 monomers \cite{Olson}.
Usually, the self-association takes place either
through the 3-dimensional domain swapping (molecular exchanges of
structural elements in one chain with the corresponding parts of another) or by
various two-residue interactions including hydrophobic, chemical cross-linking,
and electrostatic that do not pertain to the full domains \cite{crys5aur}.
Out of about 900 cases of entanglement identified by us,
most are shallowly entangled in which the entanglement disappears
if one cuts away at most 10 residues from one or both termini.
Otherwise, the chains are considered to be entangled deeply.
We classify these entanglements into two types. In type I,
the presence of the entanglement depends on the  pulling direction.
For example, the electron transport protein PDB:5AUR, 
shown in  Fig.~\ref{examp5aur},  is considered entangled if
 one anchors N' and C' in one chain (chain C shown in blue)
and pulls by N and C in another (chain A shown in red) away
from the first chain. However, 
it becomes disentangled  if one anchors N and C and 
pulls away N' and C'.
The entanglement is considered to be of type II if its presence does not depend
on the pulling direction of the chains.
In Fig.~\ref{selectpdb}, the top two examples are of type I and the
bottom two of type II.

\vspace*{0.5cm}

The types of the entanglement are related to $L_k$.  
Type I is associated with $L_k$ of 1 or -1. Type II requires that the
absolute value of $L_k$ must be larger than 1 (as in the case of 1C1G), 
or that it switches between 1 and 2 depending on how one closes the 
open protein chains into loops. One example of the latter situation
is 4AAI (see Fig.~S1 in SM), in which one chain winds around the other 
more than once but less than twice.
 If one selects the pulling direction statistically (for example
by randomly sampling the angle with respect to the line that connects the
termini), the average value of $|L_k|$ of type I 
would between 0 and 1, because for each choice it can be only equal to either 1 or 0 
($L_k=0$ if there is no entanglement). For type II entanglement, $|L_k|$ is larger than 1 on average. 
Actually, the closure of 
the chains is decided by the pulling direction in our model,
because $L_k$ is best estimated after the chains 
are fully  pulled.
As shown in the right panels of Fig.~\ref{selectpdb}, the closure of  
fully  pulled chains is very straightforward -- we connect N with C and 
N' with C' by straight lines and does not need any ensemble
of possible closures as done in ref.~\cite{Baiesi}.
We point out that
the role of $L_k$ here is not to detect the entanglement, as
this is accomplished by pulling, but to determine the entanglement type.
We can conclude that $L_k$ depends on the pulling direction of chains 
and much less so on the specifics of the closure.

\vspace*{0.5cm}

For a more detailed discussion, we have selected 15 deeply entangled
chains. They are listed in  Table~\ref{deeppairs} together with some
structural information. Their native and  pulled conformations
are shown in Fig.~\ref{selectpdb} and in Figs.~S2 and
S3 in SM. The remaining
examples are listed in Table~SI in SM.
Among the 15 cases, 1A73, 1AV1, 4ANG, 2A68, 2A8C, 4A9Z, 5AUR, and 5FCJ
are endowed with  entanglement of type I whereas
1C1G, 2ADL, 1C4D, 2AHR, 3A1M, 3AQJ and 4AAI of type II.

\vspace*{0.5cm}

\begin{figure}[h]
\centering
\includegraphics[width=0.5\textwidth]{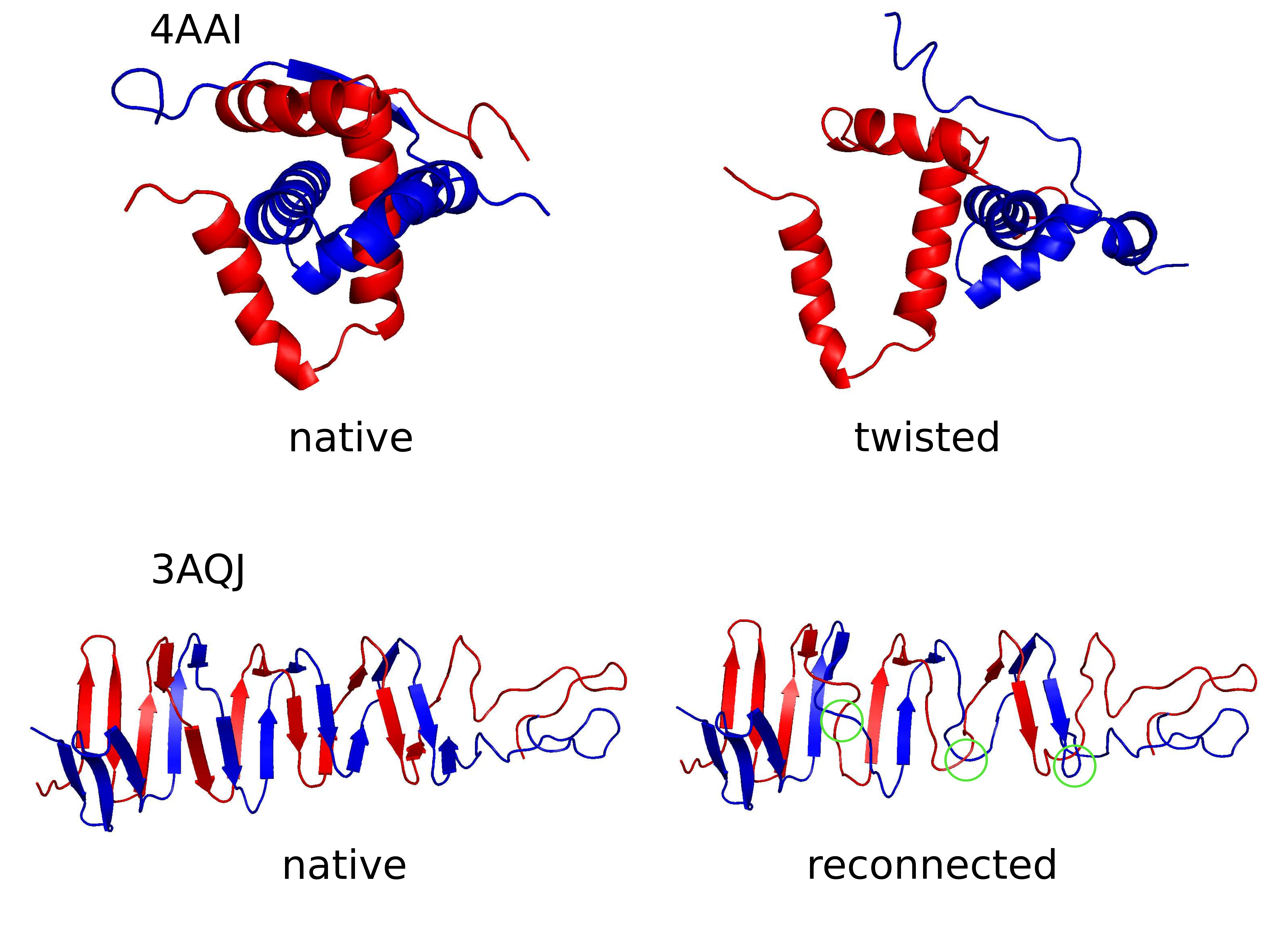}
\caption{The native (left) and twisted (right) structures of 4AAI (top), and the native (left) and reconnected (right) structures of 3AQJ (bottom). 
The green circles show the three sites in 3AQJ where the reconnection is made.}
\label{twisted}
\end{figure}

\begin{figure}[h]
\centering
\includegraphics[width=0.7\textwidth]{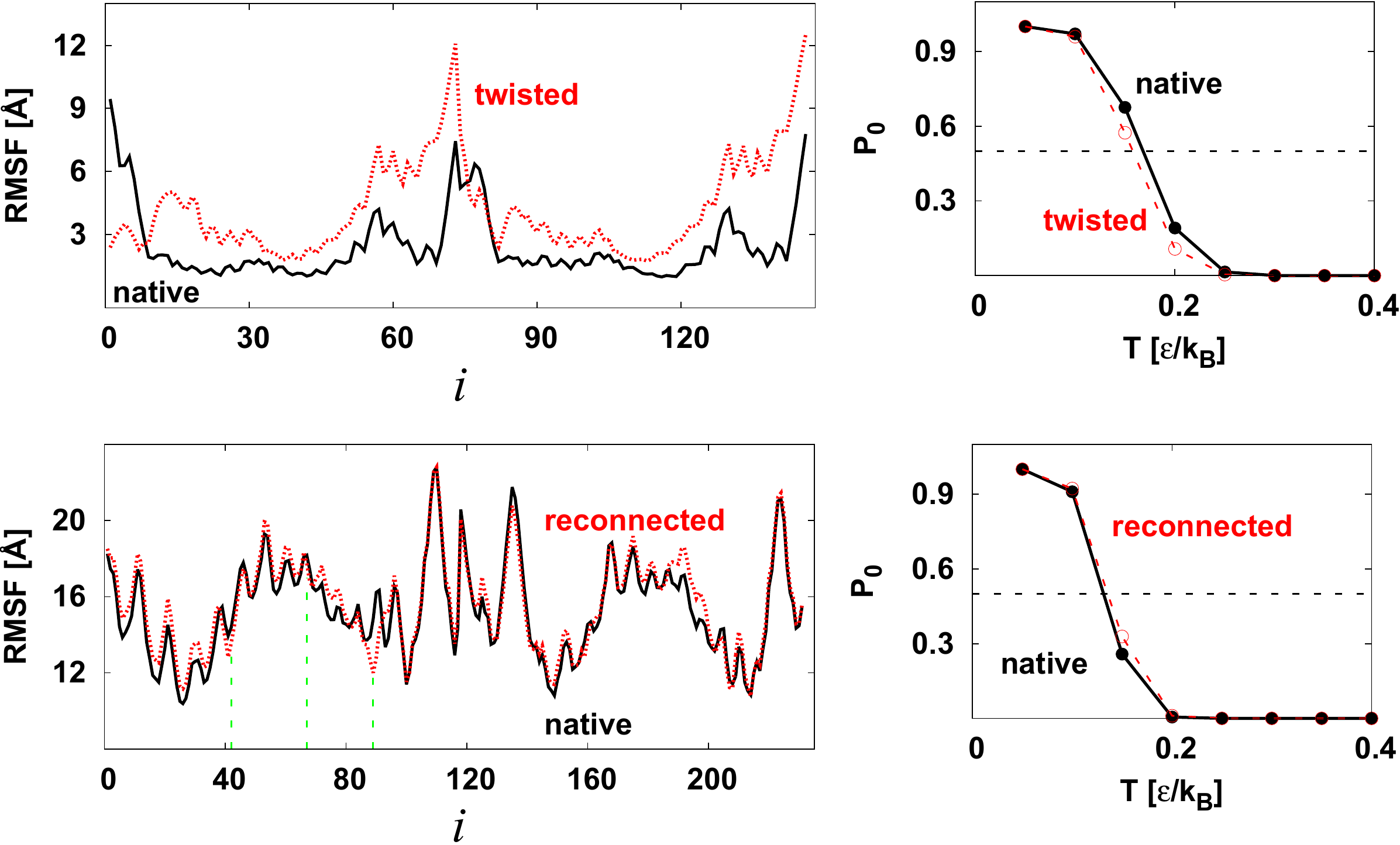}
\caption{Comparison of the physical properties corresponding to the
two forms of 4AAI (top) and 3AQJ (bottom) shown in Fig.~\ref{twisted}. The left panel shows the
RMSF at $T=0.3\; \epsilon/k_B$ and the right panel -- the $T$-dependence of $P_0$. The $T$ at which
$P_0$ crosses $\frac{1}{2}$ defines the folding temperature. 
The size of the data points is a measure of the error bars.
Chain A (B) of 4AAI ranges from residue 1 to 73 (74 to 146), while chain A (B) 
of 3AQJ ranges from residue 1 to 117 (118 to 231). The green dash lines indicate
the sites in chain A of 3AQJ at which the reconnection was made.
}\label{twisted1}
\end{figure}

The self-association of chains is expected to increase the thermal stability
of the combined system. We demonstrate it here for protein 4AAI (E73). Its two
sequentially identical chains form a large cavity on the surface of
the protein. This cavity is distal to the canonical DNA binding site
and it is positively charged. It also forms a potential ligand
binding site \cite{crys4aai}.
Fig.~\ref{twisted} (top panel) shows the entangled native structure of 4AAI on the left
and its twisted untangled version on the right. The twisted structure
is prepared by using the sculpting tool of the PyMol package \cite{pymol}.
As a result of the twist, the intrachain contacts remain the same but the
number of the interfacial contacts is reduced by 19 (the total number
of contacts in the native structure is 328). Fig.~\ref{twisted1} (top panel) shows
that the twist results in larger RMSF equilibrium fluctuations
in almost the entire chain and the downward shift by about 0.017~$\epsilon /k_B$, 
(i.e. by about 14 K) in the melting temperature $T_0$ defined as a $T$ at which
the equilibrium-calculated $P_0$ crosses $\frac{1}{2}$. 
$P_0$ is defined as the probability that all native contacts are present
simultaneously -- for a discussion of this concept see ref\cite{Wolek}.

\vspace*{0.5cm}

Another example of a situation in which the role of the entanglement
can be assessed is the 3-chain protein 3AQJ
(the bacteriophage P2 tail spike protein) \cite{crys3aqj}.
The chains of 3AQJ  wind around a threefold symmetry
axis to form a triangular pyramid and the interior of the pyramid
is occupied by hydrophobic side chains that provide stabilization.
Here, we consider the entanglement of chains A and B that have
$L_k$ of 3.
In this case, the entanglement can be removed by cutting the backbone
at the three sites indicated in the bottom panels of Fig.~\ref{twisted}
and then reconnecting on the other side of the intersection by
using the PyMol package.
As a result of
this procedure, the number of the interfacial contacts in the
reconnected structure is increased by 19 because the chains
get closer together.  This is opposite to
the effect of twisting applied to 4AAI.
At the same time, the intrachain
contacts remain the same (there are 387 native contacts
in the original native structure). These additional contacts slightly
increase $T_0$ from 0.137 to 0.141 $\epsilon/k_B$.
Despite the bigger number of the interchain contacts,
the reconnected structure is less stable when assessed based
on the values of RMSF obtained at $T=0.3\;\epsilon /k_B$
(the left bottom panel of Fig.~\ref{twisted}).
The RMSF for the reconnected structure is larger than for
the native structure at almost all sites.
Notice that $P_0$ is a measure of how many native contacts are present
independent of what is the actual distance between the corresponding
$\alpha$-C atoms, as long as the distance is below a threshold,
whereas RMSF is a measure of the positional fluctuations.
Thus the RMSF is a better characteristic of the structural
stability in this context and we conclude that the entanglement
indeed enhances the stability.

\vspace*{0.5cm}

The entanglement of chains enhances the thermal stability, but is
also expected to affect the very functioning of the complexes.
Here, we just list some examples of $n$-mers in which the role of the
entanglement should be elucidated.
The tetrameric 1AV1 (truncated
human apolipoprotein A-I) forms an antiparallel four-helix bundle shaped into an
elliptical ring that keeps the hydrophobic residues at the interface \cite{crys1av1}.
The tube-like dimeric
1C4D (gramicidin) forms an ion channel, which facilitates transport of ions
across the hydrophobic barrier of the phospholipid membranes \cite{crys1c4d}.
The dimeric 4ANG
(PRR1 coat protein) enhances
the exposed binding surface for the RNA \cite{crys4ang}.
The homodimeric 1A73 (homing endonuclease) promotes deformation
of the homing-site DNA that facilitates cleavage across the minor
groove of DNA \cite{crys1a73}.
The coiled coil dimeric 1C1G (tropomysin) stabilizes association
of actin with non-muscle cells \cite{crys1c1g}.
These dimers wrap around actin filaments and form a head-to-tail overlapping
polymer. This structure allows for flexibility between the dimers
and makes tropomyosin stay strain-free along the filament.

\section{Folding and stretching of the knotted dimeric proteins}

\begin{figure}[h]
\centering
\includegraphics[width=0.7\textwidth]{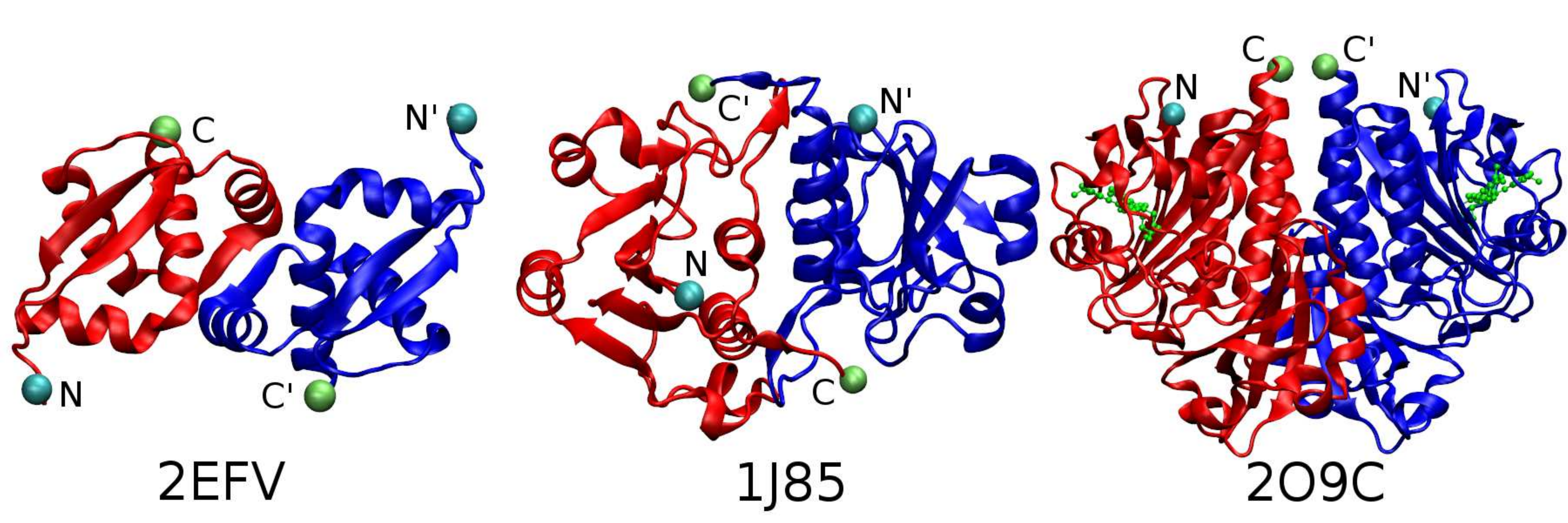}
\caption{Structural representations of the proteins studied here 2EFV (left),
1J85 (middle) and 2O9C (right). Monomers A and B of these homodimeric
structures are shown in red and blue respectively. The N and N' termini
are marked in cyan and the C and C' termini in lime.
The ligand of 2O9C is colored in green. }\label{terminus}
\end{figure}

We now consider the three homodimeric proteins 2EFV, 1J85, and 2O9C
that have been mentioned in the Introduction. Their structures are
shown in Fig.~\ref{terminus}. Folding is
considered as a function of $T$ and is based on between 200 and 500 trajectories.
Stretching is considered at $T=0.3\;\epsilon/k_B$
and is based on 50 trajectories that last for at least 200 000 $\tau$
to ensure achieving separation if it takes place.
 Stretching is performed at the speed $v_p$ of 0.005 {\AA}/$\tau$.

\subsection{2EFV}
The monomer of 2EFV comprises 87 residues but the atomic coordinates of the
first five of them are not available in the structure file. The secondary
structures of 2EFV consists of four helices (23-32, 41-49, 62-71, 74-86),
two 3-10 helices (33-35, 72-73), and two $\beta$-strands (12-17, 54-59).
The knot is of the trefoil type and its ends are located at sites 11 and 73.
The sequential separation, $\Delta n$, between the knot ends is thus 63.
On stretching, $\Delta n$ goes down in stages to about
10 residues \cite{Sulkowska_2008,Israel}, similar to what one gets for the
deeply knotted 1J85 which also contains the trefoil knot \cite{Sulkowska_2008,Israel}.
For the monomeric 2EFV, it has been shown \cite{shallow} that
a) a denatured chain folds to the knotted and globular native state
much easier than the deeply knotted monomers,
b) the set of possible topological folding pathways is richer than
the one considered in ref. \cite{Micheletti} -- in addition to
trajectories with a single knot-loop there are also trajectories
with two smaller knot-loops,
c) there is a range of optimal $T$s in which knotting is successful,
d) nascent conditions boost the peak success rate to fold from 76\% to 81\%
and eliminate single-loop trajectories. The nascent conditions occur on the
ribosome \cite{Dobson,Bustamante,ribosome,ribosome1,ribosome2,Elcock}.
In the simplest description, we have modelled them as a chain growing, one
residue at a time, from a repulsive plane \cite{nascent}.

\vspace*{0.5cm}

The folding process of the dimeric 2EFV is studied at two temperatures $T=0.3$
and 0.45~$\epsilon/k_B$ to explore the role of the temperature effects.
Folding is decided based on "all" native contacts being established for the
first time. There are several possible meanings of the word "all" here and
specifically, we consider three situations: monomeric (native contacts in the
monomer), dimeric (all native contacts in the dimer), and monomeric (native
contacts of the monomer) but in a process in which the other monomer is
present dynamically -- we just do not monitor its contacts. This last case
is denoted as 'chain A (B present)'.  
Our results are displayed in Table~\ref{f2efvd}. It shows
the probability of successful knotted folding, $S_f$, the probability
of misfolding, $S_{mf}$,  (the native contacts are established but the knot
is not made or the knot is different than the one in the native state, and 
the knotting probability $S_k$. 
$S_k$ is determined independent of whether the native contacts are
established and by including knots of non-native types, if any.
The types of knots are determined visually.
The table also shows the median folding time, $t_f$, as obtained through the
determination of $S_f$, and median knotting time, $t_k$.

\vspace*{0.5cm}

For the monomeric 2EFV, $S_f$ at $T=0.45 \;\epsilon/k_B$ is much higher than
at $T=0.3 \; \epsilon/k_B$, which agrees with the previous work \cite{shallow} that
$T=0.45 \; \epsilon/k_B$ is the optimal off-ribosome folding temperature for 2EFV.
Moreover, $S_f$ of chain A is significantly increased in the presence of chain B
at $T=0.3 \; \epsilon/k_B$, while that at $T=0.45 \; \epsilon/k_B$ remains almost the same.
The interfacial interactions between the monomers seem promoting the folding of
single chains at the lower temperature.
We also found that the folding pathway of monomeric 2EFV at $T=0.3\; \epsilon /k_B$
is different than at $0.45 \;\epsilon /k_B$ as evidenced by the values of $t_f$.
At $T=0.3 \; \epsilon/k_B$, the native contacts are established about four times
faster than the time needed to make the knot. However, the opposite takes place at
$T=0.45 \; \epsilon/k_B$. Our data suggests that tying a knot is more difficult
when the effect of the thermal fluctuations is weaker.
It appears that formation of the knot at $T=0.45 \; \epsilon /k_B$
stabilizes the protein and facilitates the establishment of the contacts.

\vspace*{0.5cm}

Folding of dimers is much harder compared to monomers.
For example, $S_f$ for the dimeric 2EFV is 3\% and 6\%
at $T=0.3$ or $0.45\; \epsilon/k_B$ respectively compared to 5\% and 65\%
obtained in the monomeric case.
The successful folding of a dimer requires the folding of its two monomers
simultaneously. When one monomer is folded the other may still be fluctuating.
We observe that in 33\% trajectories at $T=0.3\; \epsilon/k_B$ (or 34\%
at $T=0.45 \;\epsilon/k_B$), chains A and B are well separated at the end
of the simulations -- the interfacial contacts are not formed.

\vspace*{0.5cm}

\begin{figure}[h]
\centering
\includegraphics[width=0.7\textwidth]{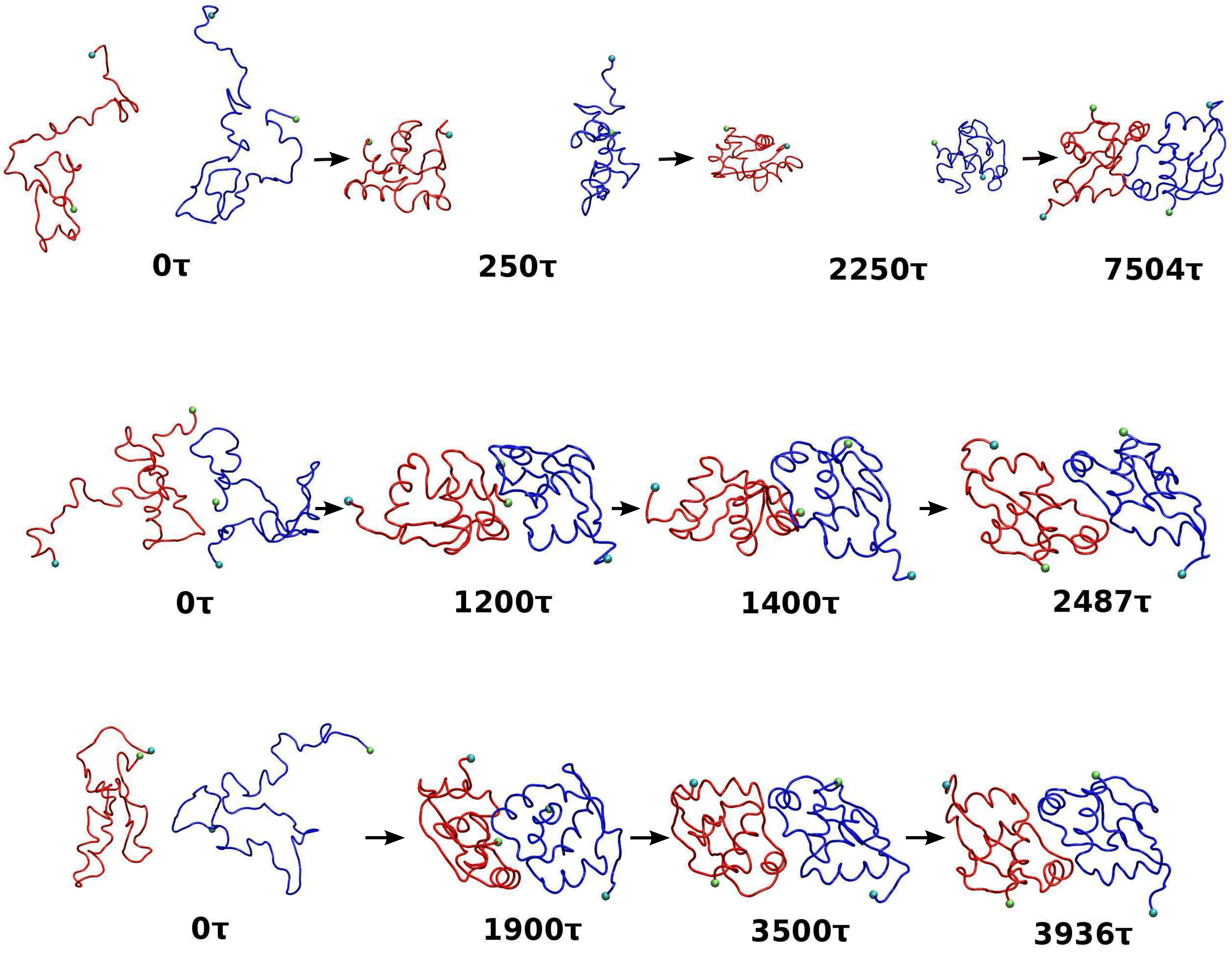}
\caption{Examples of folding pathways for the dimeric 2EFV at $T=0.3\; \epsilon/k_B$.  For $\mathbb{F}_1$ (top), the folding order is: monomer B--monomer A--interface. Monomer A of the trajectory is knotted at 6100 $\tau$, and monomer B is knotted at 5700 $\tau$. In $\mathbb{F}_2$ (middle),  the folding order is: monomer B--interface--monomer A. Monomer A of the trajectory is knotted at 1700 $\tau$, and monomer B is knotted at 1300 $\tau$.  The folding order in pathway $\mathbb{F}_3$ (bottom)  is: interface--monomer B--monomer A. Monomer A of the trajectory is knotted at 3300 $\tau$, and monomer B is knotted at 3200 $\tau$.
} \label{pathexamp}
\end{figure}

\begin{figure}[h]
\centering
\includegraphics[width=0.7\textwidth]{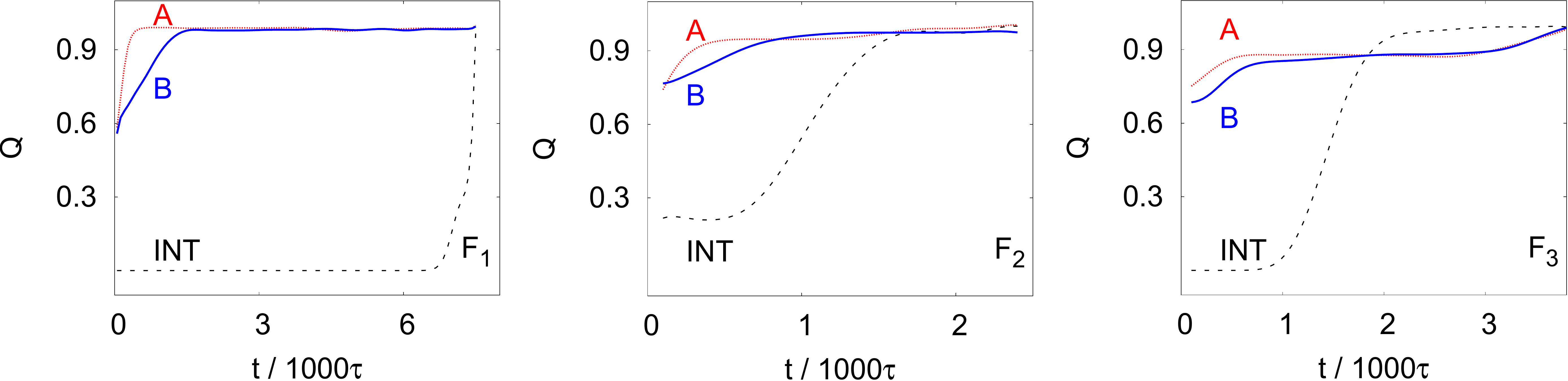}
\caption{Time evolution of $Q$ on the folding pathway $\mathbb{F}_1$, $\mathbb{F}_2$ and $\mathbb{F}_3$ of dimer 2EFV at $T=0.3\; \epsilon/k_B$.
} \label{timeQ}
\end{figure}

There are three possible pathways to fold a dimeric 2EFV starting from the fully
unfolded conformation. In the first pathway, denoted as $\mathbb{F}_1$, two monomers fold
 nearly simultaneously, which is followed by the establishment of the interfacial contacts.
In the second pathway, $\mathbb{F}_2$, one monomer folds first and then the interfacial
contacts are established, which is followed by folding of the other monomer.
In the third pathway, $\mathbb{F}_3$, interfacial contacts are established the first,
which is followed by a simultaneous folding of the monomers.
The results on the probabilities and time scales corresponding to the
three pathways are listed in Table~\ref{dimerpath}.
Examples of the folding pathways at $T=0.3\; \epsilon/k_B$ can be found
in Fig.~\ref{pathexamp}. The evolution patterns can be characterized by
providing the average fractions of the established native contacts $Q_A$, $Q_B$
and $Q_{INT}$ in monomers A, B, and at the interface respectively.
Symbol $Q$ will denote the fraction of all native contacts that are established.
The time-dependence of these parameters can be found in Fig.~\ref{timeQ}.
For example in pathway $\mathbb{F}_1$, $Q_{INT}$ reaches 1 in a slower way
than $Q_A$ and $Q_B$, while the opposite holds true in pathway $\mathbb{F}_3$.

\vspace*{0.5cm}

The knotting mechanisms of monomers in the dimeric structure are analogous to
those found for the monomeric 2EFV: they are of the two- and single-loop kind.
In 81\% trajectories at $T=0.3\; \epsilon/k_B$, both monomers get tied via the
two-loops mechanism. In the remaining trajectories, one monomer is tied via
the two-loops mechanism and the other by the single-loop mechanism.
For $T=0.45\; \epsilon/k_B$, both monomers get tied via the two-loops mechanism
in 47\% trajectories, and the mixture of single-and two-loop processes
is observed in the remaining trajectories.
We find that $S_k$ is larger than $S_f$  both for the monomers and dimers
(Table~\ref{f2efvd}), and both parameters increase as $T$ changes from 0.3 to
$0.45 \;\epsilon/k_B$ while the probability to misfolding goes down significantly.
 In both subcases ("chain A (B present)" and "dimer" in 
Table~\ref{f2efvd})
 of the dimeric systems, $S_{mf}$ actually disappears, indicating
that the effective crowding environment helps in tying the knots in
the individual chains.
This behavior is due to the fact that, the effectively
crowded environment limits the movement of chains, and thus reduces the 
entropy of the system.
We observe that knots tying temporarily during folding are often
different than the native one. Among these, the figure-eight 
knot is the most popular.
It arises from an improper order in which the native contacts form
(see Fig.~S4 in SM). Backtracking events are necessary to fold the
protein correctly \cite{backtrack}.  These events are more likely to occur at
$T=0.45\; \epsilon/k_B$ compared to at $T=0.3\; \epsilon/k_B$.
Consequently, the misfolding rate $S_{mf}$ of 2EFV at $T=0.3 \; \epsilon/k_B$ is
much higher than that at $T=0.45\; \epsilon/k_B$.

\vspace*{0.5cm}

We now consider stretching at $T=0.3\; \epsilon/k_B$. 
In the case of the monomeric 2EFV,
and of all other monomeric systems, we stretch by N and C.
We terminate stretching when the knot is tightened maximally or if the measured
force, $F$, is larger than 10~$\epsilon$/{\AA}.
Table~\ref{knotsites} shows the possible final locations of the knot ends.
The tightest trefoil knot of 2EFV has 10 residues, which agrees with the finding of
 ref.~\cite{Sulkowska_2008}.
The tightened 2EFV knot contains an entire $\alpha-$helix, such as the cases of
\{23, 32,\}, \{40, 49\},  and \{62, 71\} shown in Table~\ref{knotsites}. It may
also encompass two different secondary structures, as for location \{56,65\}.
In this case, residue 56 belongs to a $\beta$
sheet and 65 belongs to an $\alpha$ helix.
We also observe the transient location \{19,31\} in which one knot end is
within a helix and another within a turn.
Among all possible locations of the tightest knot, \{40,49\} is the most likely
-- it minimizes the walking distance from  the native location of the knot.

\vspace*{0.5cm}

For the homodimeric 2EFV (Fig.~\ref{terminus}) there are only four distinct choices
of the direction of  stretching: N--N', N--C', C--C' and N--C. The first three of these
involve the two chains in a direct way. In this case,
the protein first ruptures near the  stretched termini, then rotates so
that the dimeric interface becomes perpendicular to the direction of stretching,
and finally the two monomers get separated.
\cite{Sikoramdom}.

\vspace*{0.5cm}

\begin{figure}[h]
\centering
\includegraphics[width=0.7\textwidth]{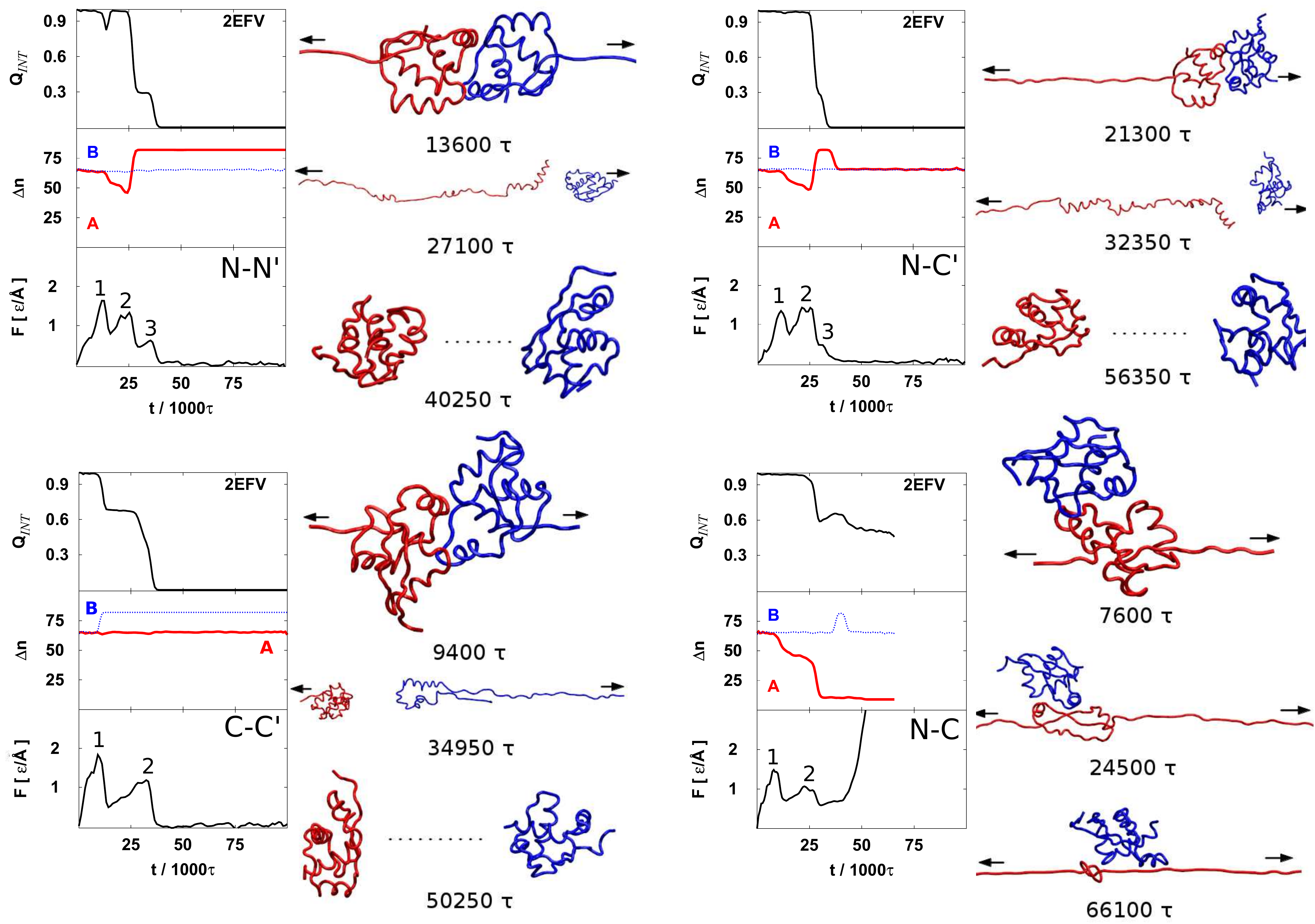}
\caption{The time evolution of $Q_{INT}$, $\Delta n$ and $F$ for the 2EFV dimer
for various schemes of  stretching, as indicated.
Examples of the conformations seen during stretching (at $T=0.3\; \epsilon/k_B$)
are shown to the right of the data panels. Chain A and B are colored in red and blue, respectively. }\label{d2efv}
\end{figure}

The force $F$, $\Delta n$ and $Q_{INT}$ as a function of time $t$ are shown
in Fig.~\ref{d2efv}. The figure also shows examples of the corresponding snapshots.
Several force peaks in the $F-t$ curve are observed in all schemes;
they are labelled by consecutive numbers in the figure. They are all
due to the tensile strain \cite{Sikoramdom} -- 2EFV has no contacting parallel
(or antiparallel) $\beta$-strands or $\alpha$-helices  to generate shear mechanical clamps.
The last force peak is always due to the separation of the two chains.
In scheme N--N', peak 1 of $1.9 \;\epsilon$/{\AA} is due to a tensile 
rupture of 6 contacts (23--58, 26--56, 26--57, 26--58, 27--58, and 27--59) 
between the backbone knot-loop and the N-terminal domain of chain A.
The N-terminal segment gets extended at this force peak.
Peak 2 of 1.7~$\epsilon$/{\AA} is due to the rupture of another
group of six contacts (31--64, 31--67, 33--70, 33--71, 34--67 and 34--70)
between the knot-loop and the N-terminal domain of chain A.
At this stage, the knot on chain A gets tightened.
The final peak has a height of 0.8~$\epsilon$/{\AA}.
A similar pattern is observed for scheme N--C'.
In scheme C--C', one chain gets untied and another remains knotted.
In the bottom left panel of Fig.~\ref{d2efv}, it is chain B that gets untied.
In this case, the first peak of 2.2~$\epsilon$/{\AA}
corresponds to dragging of the C-terminus of chain B through its
knot-loop in the process of the knot dissolution.
The second (and final)  peak has a height of 1.5~$\epsilon$/{\AA},
and is due to the separation of the two chains.

\vspace*{0.5cm}
The tensile force in schemes N--N', N--C' and C--C' results
either in the knot tightening
(see $\Delta n$ of schemes N-N' and N-C' in Fig.~\ref{d2efv}) or
in knot untying (see $\Delta n$ of scheme C-C' in Fig.~\ref{d2efv})
in single chains. Knot untying  shows as $\Delta n$ becoming equal
to the chain length of 82 (without counting the first 5 missing residues).
After the separation, both monomers relax and  the unknotted chains
can fold back to the knotted conformation or misfold, whereas the knotted chains
either untie and fold back or make the knot expand and move to another position.
Untying of the knot is common. This happens in 82\%,
100\% and 98\% trajectories in schemes N--N' and N--C' and C--C', respectively.
Our data shows that the untied monomers
separated in schemes N--N', N--C' and C--C' fold back correctly
with the probabilities of 17\%, 26\% and 24\% respectively.

\vspace*{0.5cm}

In scheme N--C, the knot on the  stretched monomer gets tightened
while the knot on the other monomer may untie temporarily through the action
of the interfacial contacts (see Fig.~\ref{d2efv}).
The simulation ends when chain A is tightened maximally or if the
force exceeds 10~$\epsilon$/{\AA} (this criterion also applies to the
other dimeric knotted proteins).
The two peaks of 1.6 and 1.3~$\epsilon$/{\AA} located at 10.0 and 22.2~$/1000\tau$
respectively (see the bottom right panel of Fig.~\ref{d2efv}) are due to
the tensile rupture of the two groups of contacts between the backbone 
knot-loop and the N-terminal domain as discussed in the context of scheme N--N'.
These processes result in tightening of the chain-A knot.
The final locations of the tightened knot are shifted compared to the locations
found for the monomeric 2EFV.
For instance, the most likely final location of monomeric 2EFV is \{40,49\},
whereas in the dimeric context it is \{44,53\}.
The location \{44,53\} is not observed when stretching monomers.

\subsection{1J85}

Folding of the deeply knotted and monomeric 1J85 has been shown to
be difficult even in structure-based models  and if it happens,
then it proceeds through a slipknot conformation \cite{dodging,Takada}.
Folding is facilitated by nascent conditions and by inclusion
of certain contacts which should count as native \cite{nascent}.
We expect that folding the dimer is even harder and we do not
attempt to study it here.

\vspace*{0.5cm}

We now focus on stretching.
Table~\ref{knotsites} shows the locations of knot ends of monomeric 1J85 under
stretching.   
The tightest knot of 1J85 contains 10 residues
\cite{Sulkowska_2008} -- as many as in 2EFV. The ends of the tightest knots are
located at the sharp turns that are seen in the native structure.
For example, location \{84,93\} contains an entire
3/10 helix, and \{94,103\} includes an entire $\beta$ sheet--$\beta$5  (see Fig.~S5 in SM). The knot ends can also be stuck
in different secondary structures such as in the case of \{69,78\} and \{78,87\}.
Here, the site 69 is inside an $\alpha$, 78 is in a $\beta$
sheet, while 87 is in an 3/10 helix. The most probable location is \{78,87\}. In this case,
the N-terminal end of the knot stays at its native position, while the
C-terminal end slides from 119 to 87. Unlike 2EFV, we observe no metastable locations for 1J85.
This may be related to the fact that in 2EFV, the native extension of the knot
is 63 sites which is substantially longer than 42 sites in 1J85.

\vspace*{0.5cm}

\begin{figure}[h]
\centering
\includegraphics[width=0.7\textwidth]{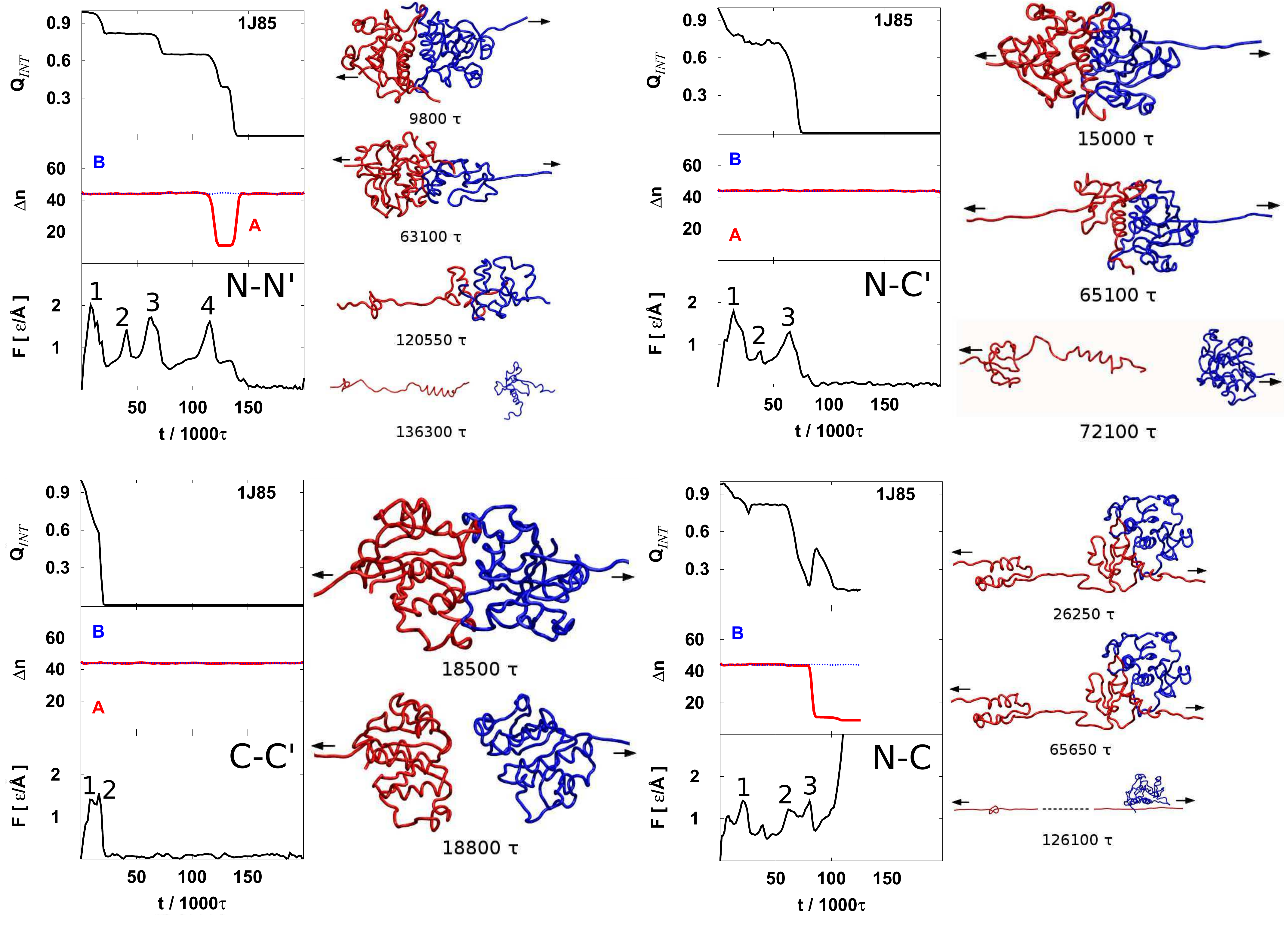}
\caption{The time evolution of $Q_{INT}$, $\Delta n$ and $F$ for the  1J85
for various schemes of  stretching as indicated.
Conformations corresponding to characteristic stages of stretching
are shown to the right of the data panels. Chain A and B are colored in red and blue, respectively. }
\label{d1j85}
\end{figure}

Again, there are four possible stretching schemes and the corresponding
time dependencies of $F$ are shown in Fig.~\ref{d1j85}. For schemes
N--N', N--C', and C--C', the maximum unravelling force, $F_{\text{max}}$,
is found to be 2.5, 2.4 and 2.3~$\epsilon$/{\AA} respectively.
The unfolding mechanism is similar to that of 2EFV. The process consists
of three steps and the separation takes place through a tensile action.
The separation involves 91 interfacial contacts.
The difference is that we observe no untying phenomena and shear forces
are involved in the first stages of stretching.

\vspace*{0.5cm}

The N--C  stretching is very similar to the monomeric case: the most likely
locations of the tightened knot are the same, but they come with different
probabilities (see Table~\ref{knotsites}). However, the least likely location
of \{69,78\} switches to \{103,112\}. The process does not affect the location
of the knot in the other chain since all 91 interfacial contacts do not involve
the residues in the native knotting core (see Fig.~S6 in SM).
In the $F-t$ curve of scheme N--C, peak 1 of 1.4~$\epsilon$/{\AA} results from
shear between $\beta 1$, $\beta 2$ and $\beta 5$.
Peak 2 is due to shear between $\alpha 1$ and $\alpha 5$, and peak 3
due to shear between $\beta 5$, $\beta 4$ and $\beta 6$ of chain A.
The conformation of chain B does not change much.

\vspace*{0.5cm}

In scheme N--N',
peak 1 is due to shear between the parallel $\beta$-sheets $\beta 1$,
$\beta 2$ and $\beta 5$ (see Fig.~S5 in SM) of chain B,
while peak 2 is due to shear between $\beta 5$, $\beta 4$ and $\beta 6$
in the same chain. Chain A unfolds after chain B. Peak 3 is due to shear
between $\beta 1$, $\beta 2$ and $\beta 5$ of chain A, and peak 4 between
 $\beta 5$, $\beta 4$ and $\beta 6$.
In scheme N--C', the peaks 1 and 3 are due to sheared $\beta 1$, $\beta 2$
and $\beta 5$ in chain A and then $\beta 5$, $\beta 4$ and $\beta 6$
in the same chain. Peak 2 is due to shear between $\alpha 1$ and $\alpha 5$
in chain B.
For scheme C--C', the two close peaks in the $F-t$ curve result from
shear between $\alpha 1$ and $\alpha 5$ of chain A (first peak) and B (second peak).

\subsection{2O9C}

Protein 2O9C contains a deep figure-eight knot.
The crystal structure of the dimerized
complex 2O9C ~\cite{crys2o9c} indicates existence of 51 interfacial contacts.
Each chain has 322 residues (the first three of them are not provided in
the PDB file). The knot ends are located at residues 30 and 271 so
the length of the knotting core is 242. The core contains a covalently bound
ligand LBV (see Fig.~\ref{terminus}).  
The C3$^2$ carbon of LBV is bound
to the $24$th residue of 2O9C.

\vspace*{0.5cm}

We first consider folding of the monomeric system -- a process that appears
not to be studied experimentally. Fig. ~\ref{foldmono2o9c} shows that
the attachment of the ligand generally does not affect the success
in folding except that at the optimal temperature of 0.3~$\epsilon /k_B$:
the ligand boosts the effectiveness of the process from 10.5 to 13\%.
In all successfully
folded trajectories at the optimal $T$, folding proceeds by first making
the knot and then by establishing the native contacts.
This is similar to what is observed in simulations for 1J85 but only at
unrealistically high $T$s \cite{Scientific}.
Examples of the folding procedure of holo and apo 2O9C at the optimal folding
$T$ are shown in Fig.~S7 in SM.
The success in proper folding for 2O9C is higher than for 1J85 but lower
than for 2EFV. This may reflect the fact that the knot in 2O9C is shallower
than in 1J85 but deeper than in 2EFV.
We did not succeed in generating folding trajectories for the dimeric 2O9C,
which is consistent with the dimer forming in the solution from two folded chains.

\vspace*{0.5cm}

Experimental stretching of the 2O9C monomer \cite{Rief} has been performed
along the direction set by sites 18 and 314, at a constant speed of 1 $\mu$m/s.
The knot was shown  to contract to $17\pm 3$ residues, as obtained by measuring
the countour-length of the protein. It has also been found that there is a
significant difference in the unfolding forces between the holo
(with the ligand) and apo (without the ligand) forms.
The apoprotein unfolds at a force of around 47 pN, while the holoprotein
unfolds at 73 pN. The higher forces in the case of holo-2O9C are needed to
unfold the ligand that is buried within the protein.
The locations of the knot ends of the monomeric 2O9C under mechanical stretching is
shown in Table~\ref{knotsites} both for the holo and apo forms.
The presence of the ligand is seen to limit the movement of the knot ends of 2O9C.
In case of holoprotein, the location of the knot ends only has three possibilities:
\{44, 122\}, \{100, 122\} and \{175, 187\}, and only 10\% of the trajectories are
stretched to the tightest knot encompassing $13$ residues (see \{175, 187\}).
For apoprotein, four more possible locations of the knot ends are observed and
72\% of them are tightened maximally (as for \{175, 187\} and \{192, 204\}).
This indicates that a smaller force is needed to stretch apo 2O9C to
arrive at the maximally tightened knot than holo 2O9C, which is consistent
the experimental observations \cite{Rief}.
However, the span of 13 residues in the tightened knot is smaller than
that of $17\pm 3$ derived experimentally. This difference either
reflects the coarse-grained character of our model, which makes the
effective size of a residue to be smaller than in an all-atom description,
or -- more likely -- that the experiment did not go to sufficiently large
forces.

\vspace*{0.5cm}

For the dimeric 2O9C, we only consider the holo form.
In the case of scheme N--C, the knot ends on chain N'--C' essentially do not move,
while the knot on chain N--C tightens in steps but never acquires the
maximally tightened shape observed in the monomeric simulations
(see Table~\ref{knotsites}).
The first force peak of 2.3~$\epsilon$/{\AA} is due to shear involved
in  stretching the ligand of chain N--C out of its pocket. The second
peak of 3.1~$\epsilon$/{\AA} is arises from shear between three paralleled 
$\beta$ sheets (shown in purple in Fig.~S5 in SM), and it is stabilized by the
immobilization of two regions in the knot-loop of chain A
(Fig.~S5 in SM, the segments in blue and green) in the process
of knot tightening. 

\vspace*{0.5cm}

\begin{figure}[h]
\centering
\includegraphics[width=0.7\textwidth]{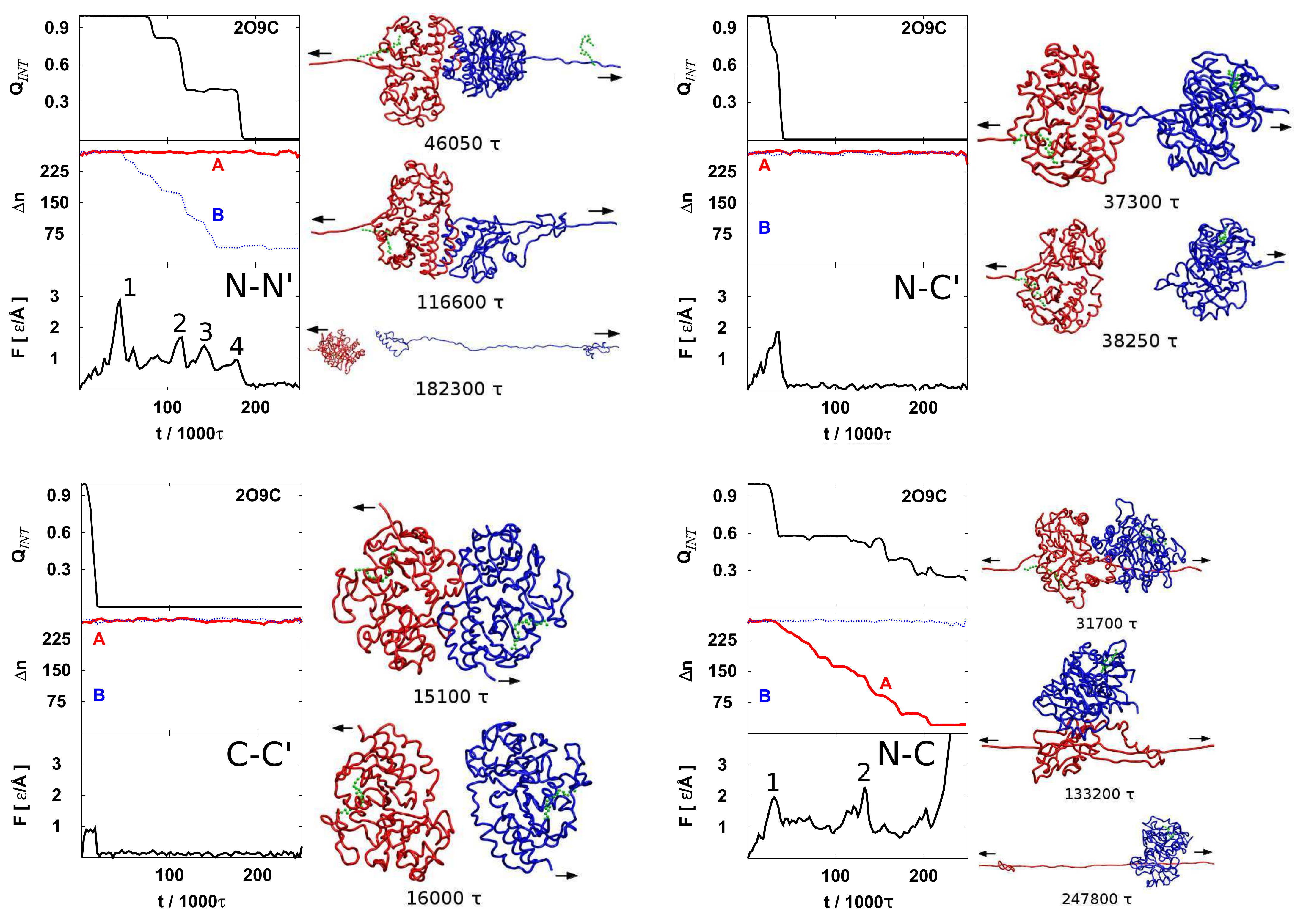}
\caption{Similar to Fig.~\ref{d1j85} but for the 2O9C dimer. The ligand of 2O9C is colored in green.
}\label{d2o9c}
\end{figure}

In schemes N--N', N--C' and C--C' the monomers get separated through the tensile
clamp (see Fig.~\ref{d2o9c}), similar to what happens in the dimeric 1J85 and 2EFV.
The force peaks are either due to shear in the interdomain contacts or due to
 tensile forces. In scheme N--N',
peak 1 of  3.4~$\epsilon$/{\AA} is due to the ligand-related shear.
Peak 2 of 2.0~$\epsilon$/{\AA}  is due to shear between
three paralleled $\alpha$ helices (as shown in Fig.~S5
in SM and marked in orange). Peak 3 of  1.7~$\epsilon$/{\AA} due to shear
between three paralleled $\beta$ sheets (green), while peak 4 of
1.3~$\epsilon$/{\AA} arises from shear between two $\beta$ sheets 271 to 279
(pink) and 282 to 291 (green) of chain B. In scheme N--C' there is only one peak.
Its height is 2.5~$\epsilon$/{\AA}. It arises from shear involved in
 stretching the ligand of chain A out of its pocket. The tensile peaks
associated with the separation of the two chains in schemes N--N' and
N--C' are hard to notice, because they almost coincide with the
preceding shear-based peak.
In scheme C--C', the peak of 1.3~$\epsilon$/{\AA} is due to the tensile forces
of separation.

\section{Conclusions}

In this paper, we have proposed a pulling-based method to define structural
entanglements. Unlike the standard stretching, the method involves
all four termini of the two chains that are tested. 
 When making surveys of multi-meric protein structures, we
have identified two types of  entanglements. In type-II entanglement,
there is no chain separation regardless of the direction of pulling
and a necessary condition for this situation is that the Gaussian linking
number is not smaller than 1.  In type I, for which $L_k$ is not 
larger than 1, the entanglement may or may not show, depending of 
how one pulls.
The entanglements
may play various roles and one of them is an enhancement in the
thermal stability of the complex. 


\vspace*{0.5cm}

\begin{figure}[h]
\centering
\includegraphics[width=0.6\textwidth]{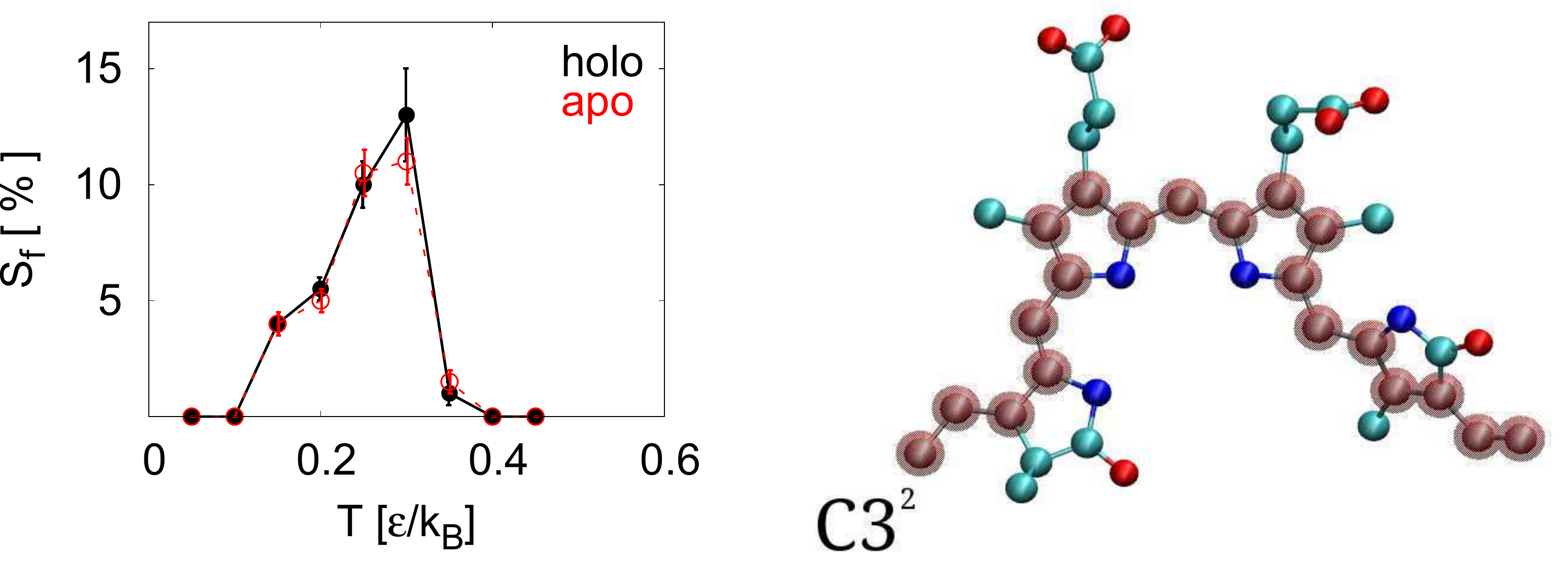}
\caption{Left: The success rate of proper folding of monomeric 2O9C in both holo
and apo forms as a function of $T$.
Right: Schematic representation of the atomic structure of ligand LBV. The carbons, nitrogens and oxygens of the ligand are displayed as cyan, blue and red beads, the hydrogen atoms are not shown. The red-cyan beads are the backbone carbon atoms of the ligand.
}\label{foldmono2o9c}
\end{figure}

We have also considered folding and stretching properties of three
homodimers 2EFV, 1J85 and 2O9C, which previously have been studied
only as single chains. We find that the equilibrium
fluctuations RMSF of one chain in the presence of its companion
are similar to those of the single chain (see Fig.~S8 in SM),
except for the region 122--130 in 1J85 where RMSF is suppressed.
In spite of this, the presence of the other monomer may result
in an enhancement in the success of knotted  folding.
Folding to the knotted state depends on the temperature at which it is
studied but the optimal folding $T$ does not depend on whether one studies
the dimeric or the corresponding monomeric system.
We find that, for monomers, the ease of folding decreases with
the depth of the native knot: the monomeric 2O9C folds easier than
the monomeric 1J85.
We have also shown that the presence of the LBV ligand in 2O9C slightly
enhances the probability of a successful folding (see Fig.~\ref{foldmono2o9c}).

\vspace*{0.5cm}

Stretching of the dimers proceeds similarly to that of the corresponding
monomers except that there is a tensile-based separation when two chains
are involved. If one chain is involved, the presence of the other chain
may affect the probabilities of acquiring the final location of the
tightened knot. It appears that the experimental studies of stretching
of 2O9C ~\cite{Rief} did not probe sufficiently large forces to achieve
the ultimate tightening of the knot.
The shallowly knotted 2EFV has a rich spectrum of possible behaviors
not only during folding but also in stretching. In the monomeric case,
the knot in 2EFV can only get tighter. However, in the dimeric case,
it may unravel in one chain but not in the other. If it unravels,
it may tie back after the separation. Such a variety is not
observed in 1J85 and 2O9C.\\

{\bf Supplementary Material.} 
The supplementary material (SM) contains a full list of entangled
protein chains that were studied. The SM also provides additional figures. They show: examples of the native and stretched conformations corresponding to type I and type II entanglements; the transient figure-eight knot that arises during folding of 2EFV; a schematic representation of protein 1J85, the crystal structure of protein 2O9C; the interfacial contacts and the RMSF for 2EFV, 1J85, and 2O9C; examples of folding pathways for holo- and apo-2O9C.

{\bf Acknowledgments}
This work has been supported by the National Science Center in Poland
under the aegis of the EU Joint Programme
in Neurodegenerative Diseases (JPND); grant number 2014/15/Z/NZ1/00037.
The local computer resources were supported by the PL-GRID network
and financed by the European Regional Development Fund
under the Operational Programme Innovative Economy NanoFun POIG.02.02.00-00-025/09.

\clearpage
\begin{table}[ht]
\caption{The information on 15 deeply entangled protein pairs.}
\centering
\begin{tabular}{|c |c |c | c|}
\hline
\multirow{1}{*}{classification} & PDB code  & list of entangled pairs &structural information\\
\hline
HYDROLASE & 1A73 & A--B & \begin{tabular}{c} X-Ray diffraction, hexameric. \\Van der Waals contacts, \\C-terminal tail is domain-swapped. \end{tabular}\\
\hline
LIPID TRANSPORT & 1AV1 & \begin{tabular}{c} A--B, A--C, A--D, \\B--C, B--D, C--D \end{tabular} & \begin{tabular}{c} X-Ray diffraction, tetrameric. \\ Hydrophobic interactions.   \end{tabular}\\
\hline
VIRUS & 4ANG & A--B & \begin{tabular}{c} X-Ray diffraction, 90 dimers. \\ Domain swapped.   \end{tabular}\\
\hline
TRANSFERASE & 2A68 & \begin{tabular}{c} A--B, D--E, K--L, \\ N--O, C--D, M--N \end{tabular} & \begin{tabular}{c} X-Ray diffraction, hexameric. \\ Polar and hydrophobic interactions.  \end{tabular}\\
\hline
LYASE & 2A8C & C--E, D--F & \begin{tabular}{c} X-Ray diffraction, tetrameric. \\ Hydrogen bonds.   \end{tabular}\\
\hline
TRANSCRIPTION & 4A9Z & \begin{tabular}{c} A--B, C--D, \\ A--C, B--D \end{tabular} & \begin{tabular}{c} X-Ray diffraction, tetrameric. \\ Unpublished results.   \end{tabular}\\
\hline
ELECTRON TRANSPORT & 5AUR & A--C, E--G & \begin{tabular}{c} X-Ray diffraction, dimeric. \\ Domain swapped.   \end{tabular}\\
\hline
RIBOSOME & 5FCJ & \begin{tabular}{c} D (bundle1)\cite{footnote}--W (bundle1)
 \\N (bundle2)--T (bundle3) \\E (bundle4)--A (bundle5) \end{tabular}&X-Ray diffraction.\\
 \hline
CONTRACTILE PROTEIN & 1C1G & A--B, C--D & \begin{tabular}{c} X-Ray diffraction, dimeric. \\ Covalent bonds, coiled coil.   \end{tabular}\\
\hline
DNA BINDING PROTEIN & 2ADL & A--B & \begin{tabular}{c} NMR, dimeric. \\ Hydrogen bonds, domain-swapped.   \end{tabular}\\
\hline
\begin{tabular}{c} ANTIBIOTIC \end{tabular} & 1C4D & A--B, C--D & \begin{tabular}{c} X-Ray diffraction, dimeric. \\ Hydrogen bonds, domain-swapped.   \end{tabular}\\
\hline
OXIDOREDUCTASE & 2AHR & A--D, B--C & \begin{tabular}{c} X-Ray diffraction, decameric. \\ Hydrophobic interaction, salt bridges, \\
domain-swapped.   \end{tabular}\\
\hline
STRUCTURAL PROTEIN & 3A1M & \begin{tabular}{c} A--B, A--C, B--C,\\ D--E, D--F, E--F,\\ A--D, B--F, C--E \end{tabular} & \begin{tabular}{c} X-Ray diffraction, monomeric. \\ Unpublished results.   \end{tabular}\\
\hline
METAL BINDING PROTEIN & 3AQJ & \begin{tabular}{c} A--B, A--C, B--C,\\ P--Q, P--R, Q--R \end{tabular} & \begin{tabular}{c} X-Ray diffraction, trimeric. \\ Hydrophobic interaction, domain swapped.   \end{tabular}\\
\hline
VIRAL PROTEIN & 4AAI & A--B & \begin{tabular}{c} NMR, dimeric. \\ Hydrophobic interaction, domain swapped.   \end{tabular}\\
\hline
\end{tabular}
\label{deeppairs}
\end{table}

\begin{table}[ht]
\caption{The folding  probability of monomeric 2EFV (labeled as 'chain A (monomer)'), chain A in the presence of chain B (labeled as 'chain A (B present)') and dimer (labeled as 'dimer') at $T=0.3$ and 0.45~$\epsilon/k_B$ is denoted as $S_f$, while their misfolding rate, the knotting probability, median folding time and median knotting time are denoted as $S_{mf}$, $S_k$, $t_f$ and $t_k$, respectively. The data is obtained based on 500 trajectories for each temperature.}
\centering
\begin{tabular}{|c |c |c |c |c |c |c|c | c| c| c|}
\hline
\multirow{2}{*}{2EFV} & \multicolumn{5}{c|}{$T=0.3\; \epsilon/k_B$} & \multicolumn{5}{c|}{$T=0.45 \;\epsilon/k_B$} \\
\cline{2-11}
& $S_f$ & $S_{mf}$ & $S_k$ & $t_f$ ($\tau$) & $t_k$ ($\tau$)& $S_f$ & $S_{mf}$ & $S_k$ & $t_f$ ($\tau$)& $t_k$ ($\tau$) \\
\hline
chain A (monomer)& 5\% & 72\%& 18\% & 6186 & 24600 & 65\% & 13\% & 87\% & 8018 & 4000\\
chain A (B present)& 44$\pm$6 \% &6$\pm$3\%& 60$\pm$1\% & 5075$\pm$1225 & 17375$\pm$4575 & 50$\pm$15\% & 0\% & 91$\pm$1\% & 290850$\pm$70625 & 4400 \\
dimer & 3\% & 19\% & 32\% & 12854 & -- & 6\% & 0\% & 83\% & 687548 &--\\
\hline
\end{tabular}
\label{f2efvd}
\end{table}

\begin{table}[ht]
\caption{The probability of  folding dimeric 2EFV via three possible pathways .}
\centering
\begin{tabular}{|c |c |c |}
\hline
\multirow{1}{*}{pathways} & {$T=0.3 \;\epsilon/k_B$} & {$T=0.45\; \epsilon/k_B$} \\
\hline
$\mathbb{F}_1$ & 56\% & 0\%\\
$\mathbb{F}_2$ & 19\% & 3\%\\
$\mathbb{F}_3$ & 25\% & 97\% \\
\hline
\end{tabular}
\label{dimerpath}
\end{table}

\begin{table}[ht]
\caption{The locations of the knot ends of 2EFV, 1J85 and 2O9C under the N--C stretching scheme at $T=0.3\; \epsilon/k_B$.}
\centering
\begin{tabular}{|c |c |c |c |c |c | c|c|}
\hline
\multicolumn{2}{|c|}{Protein} & \multicolumn{6}{|c|}{locations of the knot ends} \\
\hline
\multirow{4}{*}{2EFV} & \multirow{2}{*}{monomer} & {\{19,31\}} & {\{23,32\}} &{\{40,49\}} & {\{56,65\}} & {\{62,71\}} & --\\
& & 8\%& 8\% &66\%& 2\% &16\% &--\\
\cline{2-8}
& \multirow{2}{*}{dimer} & {\{44,53\}} & {\{38,47\}} &{\{40,49\}} & {\{56,65\}} & --&--\\
& & 52\%& 6\% &24\%& 18\% &--&--\\
\hline
\multirow{4}{*}{1J85} & \multirow{2}{*}{monomer}& {\{69,78\}} & {\{78,87\}} &{\{84,93\}} & {\{94,103\}} &--&--\\
& & 2\% &44\%&22\%&32\% & --&--\\
\cline{2-8}
 & \multirow{2}{*}{dimer} & {\{103,112\}} & {\{78,87\}} &{\{84,93\}} & {\{94,103\}} &--&--\\
 & & 6\% &24\%&34\%&36\% & --&--\\
\hline
\multirow{5}{*}{2O9C} & \multirow{1}{*}{monomer}& {\{44,122\}} & {\{100,122\}} &{\{175,187\}} & {\{187,200\}}& {\{192,204\}} & {\{204,248\} or \{204,284\}}\\
&holo & 54\% &36\%&10\%&-- &-- &--\\
&apo & 6\% &2\%&66\%&10\%& 6\% &10\%\\
\cline{2-8}
 & \multirow{1}{*}{dimer} & {\{44,122\}} & {\{100,122\}} &-- & --& -- & -- \\
 & holo& 92\% &8\%&-- & --& -- & -- \\
\hline
\end{tabular}
\label{knotsites}
\end{table}

\begin{thebibliography}{99}
\bibitem{footnote}
The data of large structures is usually split over multiple compatible pdb files, which are combined into single bundles.
\end{thebibliography}
\end{document}